\documentclass{PoS}
\usepackage{amsmath}
\usepackage{subfig}
\usepackage{cite}
\usepackage{url}
\usepackage{slashed}
\usepackage{subfig}

\newcommand{\pt}{\mathbf{p}}
\newcommand{\kt}{\mathbf{k}}
\newcommand{\qt}{\mathbf{q}}
\newcommand{\qtt}{\mathbf{\tilde{q}}}

\title{TMD splitting functions and the corresponding evolution equation}

\ShortTitle{TMD splitting functions and the corresponding evolution equation}

\author{\speaker{A.\ Kusina}\,$^{a}$$^{b}$\\
       \llap{$^a$}Laboratoire de Physique Subatomique et de Cosmologie,
                  CNRS/IN2P3,\\ 53 avenue des Martyrs, 38026 Grenoble, France\\
       \llap{$^b$}Institute of Nuclear Physics Polish Academy of Sciences,
                  PL-31342 Krakow, Poland\\
       E-mail: \email{kusina@lpsc.in2p3.fr}}

\abstract{We generalize the Catani Hautmann formalism to calculate the $P_{gq}$
and $P_{qq}$ transverse momentum dependent splitting functions.
Then we use the obtained $P_{gq}$ kernel to construct a low-$x$
evolution equation for gluons that takes into account the effect of non-diagonal
quark-to-gluon splittings. In order to write down a consistent equation we
resum virtual corrections coming from the gluon channel and demonstrate that
this implies a suitable regularization of the $P_{gq}$ singularity, corresponding to
a soft emitted quark. We also note that the obtained equation is in a straightforward
manner generalized to a nonlinear evolution equation which takes into account
effects due to the presence of high gluon densities.}

\FullConference{ 25th International Workshop on Deep Inelastic Scattering and Related Topics \\
         3-7 April 2017 \\
         University of Birmingham, Birmingham, UK}
         
\begin{document}
%====================================================================

% 5 pages for parallel talks

%%%%%%%%%%%%%%%%%%%%%%%%%%%%%%%%
\section{Introduction}
\label{sec:intro}
%%%%%%%%%%%%%%%%%%%%%%%%%%%%%%%%
Parton distribution functions (PDFs) together with parton level matrix
elements allow for a very accurate description of ``hard'' events in
hadron-hadron and hadron-electron collisions.
The bulk of such analysis is carried out within the framework of collinear
factorization~\cite{Ellis:1978sf,Collins:1984kg}.
However, there exist classes of multi-scale processes where the use of
more general schemes is of advantage.
An example of such a process is a high-energy or low $x$ limit of hard processes
with $s \gg M^2 \gg \Lambda_{\text{QCD}}^2$ where $x = M^2/s$.
In such a scenario it is necessary to resum terms enhanced by logarithms
$\ln 1/x$ to all orders in the $\alpha_s$, which is achieved by BFKL evolution
equation~\cite{Kuraev:1976ge,Balitsky:1978ic}.
The resulting formalism called high-energy or $k_T$
factorization~\cite{Catani:1990eg,Catani:1990xk} provides
a factorization of such cross-sections into a TMD coefficient or ``impact factor''
and an ``unintegrated'' gluon density.

Unfortunately, the high-energy factorization framework has important limitations
that makes it cumbersome to apply it directly to an arbitrary process. For example
it takes into account only gluon densities which is not acceptable for some quark
initiated processes. What is more, since it is valid only in the low $x$ region
($x\lesssim10^{-2}$) by construction it is restricted to exclusive observables such that
$x$ of both gluons is fixed; e.g. description of processes involving fragmentation function,
requiring integration over full $x$ range, rises problems.

In this contribution we summarize our efforts~\cite{Gituliar:2015agu,Hentschinski:2016wya}
to formulate a prescription to accommodate the advantages of both collinear and high-energy
factorizations. In particular, our goal is to construct evolution equation
(or rather a system of equations)
for unintegrated parton distributions (with $k_T$-dependence) that would have the following
properties:
(i) it should resum the low $x$ logarithms,
(ii) has smooth continuation to the large $x$ region,
(iii) include both quarks and gluons,
and (iv) reproduce the correct collinear limit given by DGLAP.
%
% To achieve this goal
To this end we extend the Catani-Hautmann (CH)~\cite{Catani:1994sq} and
Curci-Furmanski-Petronzio (CFP)~\cite{Curci:1980uw} formalisms, which allows us for calculation
of $k_T$-dependent splitting functions. For the moment we concentrate only on the real part
of the quark splitting functions, and in the second step we use the newly calculated $P_{gq}$
splitting to construct a low $x$ evolution equation incorporating quark contributions.

%%%%%%%%%%%%%%%%%%%%%%%%%%%%%%%%
\section{TMD splitting functions}
\label{sec:kernels}
\begin{figure}[!t]
\centerline{
\subfloat[]{\includegraphics[height=4.5cm]{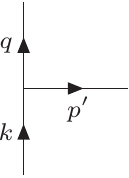}
\label{subfig:kinem}}
\qquad\qquad
\subfloat[]{\includegraphics[height=4.5cm]{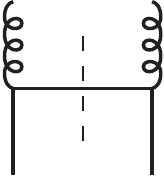}
\label{subfig:pgq}}}
\caption{(a) Kinematics of a parton splitting. An initial parton with transverse momentum
${\bf k}$ splits into a parton with transverse momentum ${\bf q}$, emitting an on-shell
parton with momentum $p'=k-q$.
(b) The only diagram contributing to determination of the first order $P_{gq}^{(0)}$
TMD splitting function.}
\label{fig:kinem}
\end{figure}

%%%%%%%%%%%%%%%%%%%%%%%%%%%%%%%%
So far we have calculated the real emission parts of the quark TMD splitting functions
($P_{qg}$, $P_{gq}$, $P_{qq}$). The method used for the calculation is based on the
two-particle-irreducible (2PI) expansion of refs.~\cite{Curci:1980uw} and~\cite{Catani:1994sq}.
The details of the method can be found in~\cite{Gituliar:2015agu} where the computations
were performed. In this contribution we only highlight two crucial ingredients used in the
calculation. The first are projection operators used to obtain factorization by decoupling the
2PI kernels in the momentum and helicity space. These projectors are generalization of the
ones introduced in~\cite{Curci:1980uw} accounting for the more general kinematics (featuring
off-shell momenta on the incoming legs of the 2PI kernels):
\begin{equation}
 \mathbb{P}_{g,\,\text{in}}^{\,\mu\nu} = \frac{k_\perp^\mu k_\perp^\nu}{\kt^2},
 \qquad
 \mathbb{P}_{q,\,\text{in}} = \frac{y \, \slashed{p}}{2},
\end{equation}
where the corresponding kinematics is shown in Fig.~\ref{subfig:kinem} and the involved momenta
are parametrized as follows:
\begin{equation}
k^\mu = y p^\mu + k_\perp^\mu,
\qquad
q^\mu = x p^\mu + q_\perp^\mu + \frac{q^2+\qt^2}{2x p\cdot n} n^\mu,
\qquad
p'=k-q.
\end{equation}
%--------
%--------
%
The second new element is the construction of an appropriate vertex that can be used in
the presence of off-shell particles such that the obtained results are gauge invariant.
We quote here only one of the vertices that is needed for calculation of the $P_{gq}$
splitting:
\begin{equation}
\Gamma_{g^*q^*q}^\mu (q, k, p') = igt^a \left(\gamma^\mu - \frac{p^\mu}{p\cdot q} \slashed{k}\right).
\end{equation}
The remaining vertices and explanations on how we construct them can be found in~\cite{Gituliar:2015agu}.

The prescription to compute the TMD splitting function $\tilde{P}_{ij}^{(0)}$ follows from
the ladder expansion of refs.~\cite{Curci:1980uw,Catani:1994sq} and reads:
\begin{equation}
\begin{split}
\hat K_{ij} \left(z, \frac{\kt^2}{\mu^2}, \epsilon \right) &=
z \int \frac{d^{2 + 2 \epsilon} {\qt}}{2(2\pi)^{4+2\epsilon}}
     \underbrace{\int d q^2 \, \mathbb{P}_{j,\,\text{in}} \otimes
                 \hat{K}_{ij}^{(0)}(q, k) \otimes \mathbb{P}_{i,\,\text{out}}}
                    _{\tilde{P}_{ij}^{(0)} \left(z, \kt, \qtt, \epsilon \right)}
     \,\Theta(\mu_F^2+q^2),
\end{split}
\label{eq:kernel}
\end{equation}
where $\hat{K}_{ij}^{(0)}$ is the first order expansion of the 2PI kernel of~\cite{Curci:1980uw},
and $\mathbb{P}_{j,\,\text{in}}$ and $\mathbb{P}_{i,\,\text{out}}$ are the projection operators
discussed above. As an example, in Fig.~\ref{subfig:pgq}, we show the diagram representing the
first order expansion of $\hat{K}_{gq}^{(0)}$ which is the only diagram necessary to compute
the NLO $P_{gq}^{(0)}$ TMD splitting function in the presented approach.
Using eq.~\eqref{eq:kernel} and integrating the angular dependence we obtain the following 
angular averaged TMD splitting functions:
\begin{align}
P_{qg}^{(0)} \left(z, \frac{\kt^2}{\qtt^2}, \epsilon \right) &= 
T_R \left(\frac{\qtt^2}{\qtt^2 + z(1-z)\,\kt^2}\right)^2
\Bigg[ z^2+(1-z)^2 + 4z^2(1-z)^2 \frac{\kt^2}{\qtt^2} \Bigg],
\\
{P}_{gq}^{(0)} \left(z, \frac{\kt^2}{\qtt^2}, \epsilon \right) &=
C_F \Bigg[ \frac{2\qtt^2}{z |\qtt^2-(1-z)^2 \kt^2|}
%    \\ &
% \hspace{2cm} 
% - \frac{\qtt^2 (\qtt^2(2-z) + \kt^2 z(1-z^2)) - \epsilon z (\qtt^2 + (1-z)^2 \kt^2)}{(\qtt^2 + z(1-z)\kt^2)^2}
%   \Bigg] 
- \frac{\qtt^2 (\qtt^2(2-z) + \kt^2 z(1-z^2))}{(\qtt^2 + z(1-z)\kt^2)^2}
% \\ & \hspace{2cm}
+ \frac{\epsilon z \qtt^2 (\qtt^2 + (1-z)^2 \kt^2)}{(\qtt^2 + z(1-z)\kt^2)^2}
\Bigg],
\\
{P}_{qq}^{(0)} \left(z, \frac{\kt^2}{\qtt^2}, \epsilon \right) & = 
C_F \left(\frac{\qtt^2}{ \qtt^2 + z(1-z)\kt^2} \right)
\bigg[ \frac{\qtt^2 + (1-z^2)\kt^2}{(1-z)|\qtt^2 - (1-z)^2 \kt^2|} 
\\\nonumber
&\hspace{1.2cm} +
\frac{z^2 \qtt^2 -z(1-z)(1 - 3z + z^2)\kt^2 + (1-z)^2\epsilon (\qtt^2 + z^2 \kt^2)}
                              {(1-z)  (\qtt^2 +    z(1-z)\kt^2)}\bigg],
% \\&\hspace{4.2cm} +
% \frac{z^2 \qtt^2 -z(1-z)(1 - 3z + z^2)\kt^2}{(1-z)(\qtt^2 + z(1-z)\kt^2)}\bigg]
\end{align}
where $\qtt = \qt - z \kt$ and $z=x/y$.
The $P_{qg}^{(0)}$ splitting function have been obtain
previously~\cite{Catani:1994sq,Ciafaloni:2005cg,Hautmann:2012sh} and we reproduce
this result. On the other hand the results for $P_{gq}^{(0)}$ and $P_{qq}^{(0)}$ are new.

%%%%%%%%%%%%%%%%%%%%%%%%%%%%%%%%
\section{Evolution equation with quarks}
\label{sec:evol}
%%%%%%%%%%%%%%%%%%%%%%%%%%%%%%%%
Our starting point is the leading order BFKL equation which describes evolution
in $\ln 1/x$  for  the dipole amplitude in the momentum space:
\begin{equation}
\label{eq:bkversion1}
{\cal F}(x,\qt^2) = 
{\cal F}^{0}(x,\qt^2)
+ \overline\alpha_s\int_x^1\frac{dz}{z} \int\frac{d^2\pt'}{\pi \pt'^2}
  \left[{\cal F}(x/z,|{\bold q}+{\pt'}|^2)-\theta(\qt^2-\pt'^2){\cal F}(x/z,\qt^2)\right]
\end{equation}
where $\overline\alpha_s=\frac{C_{A}\alpha_s}{\pi}$.
This form is particularly useful  to promote the BFKL equation to a system of
equations for quarks and gluons.
We will incorporate a contribution from quarks in the following form:
$\frac{\alpha_s}{2\pi}
\int_x^1\frac{dz}{z}\int\frac{d^2\pt'}{\pi \pt'^2}
P_{gq}(z,\pt',\qt){\cal Q}(x/z,|{\bold q}+{\pt'}|^2)$,
where ${\cal Q}(x/z,|{\bold q}+{\pt'}|^2)$ is a distribution of quarks.
Next we introduce a resolution scale $\mu$ allowing us to decompose the kernel of
the gluonic part of \eqref{eq:bkversion1} into a resolved real emission part
with $\pt'^2>\mu^2$ and the unresolved part with $\pt'^2<\mu^2$.
Additionally, since the integral over $\pt'$ in the quark contribution is divergent it needs
to be regulated. In the following we achieve this through introducing the same cut-off
$\mu$ as used for the gluonic part. Technically this is obtained through including
a theta function $\theta(\pt'^2-\mu^2)$ in the quark term:%
\footnote{Note that, in the gluon case, this additional scale is just
  for technical convenience as $1/\pt'^2$ is regularized by the virtual
  contribution, in the case of quarks, $\mu$ scale is really needed for
  regularizing the corresponding expression.}
\begin{align}
\label{eq:bkversion2}
{\cal F}(x,\qt^2)&={\cal F}^{0}(x,\qt^2)
+\overline\alpha_s\int_x^1\frac{dz}{z}\int\frac{d^2\pt'}{\pi \pt'^2}{\cal F}(x/z,|{\bold q}+{\pt'}|^2)\theta(\pt'^2-\mu^2)\\\nonumber
&+\overline\alpha_s\int_x^1\frac{dz}{z}\int\frac{d^2\pt'}{\pi \pt'^2}\big[{\cal F}(x/z,|{\bold q}+{\pt'}|^2)\theta(\mu^2- \pt'^2)-
\theta(\qt^2-\pt'^2){\cal F}(x/z,\qt^2)\big]\\\nonumber
&+\frac{\alpha_s}{2\pi}\int_x^1dz\int\frac{d^2\pt'}{\pi \pt'^2}P_{gq}(z,\pt',\qt){\cal Q}(x/z,|\pt'+\qt|^2)\theta(\pt'^2-\mu^2)\nonumber\,.
\end{align}
Equation~\eqref{eq:bkversion2} is in a suitable form to perform resummation of the
virtual and unresolved emissions by going to the Mellin space ($x\to\omega\to x$).
The detailed steps of this resummation are presented in ref.\cite{Hentschinski:2016wya}
in the following we only present the final result:
\begin{equation}
\begin{split}
 {\cal F}(x,\qt^2)
& =
\tilde{\cal F}^0(x,\qt^2)
% \\ &
+
\frac{\alpha_s}{2 \pi}
\int_x^1 \frac{dz}{z}  \int_{\mu^2} \frac{d^2 \pt'}{\pi \pt'^2} %\theta(\pt^2-\mu^2)
\Bigg[  
\\ &
% \hspace{3cm}
   \Delta_R(z, \qt^2, \mu^2)
 \bigg( 2 C_A
    {\cal F}\left(\frac{x}{z},|\qt+\pt'|^2\right)
+ C_F 
    {\cal Q}\left(\frac{x}{z},|\qt+\pt'|^2\right) \bigg)
\\ &
% \hspace{2.2cm}
-
     \int_z^1 \frac{dz_1}{z_1} \Delta_R(z_1, \qt^2, \mu^2)
 \left[ \tilde{P}'_{gq}\left(\frac{z}{z_1}, \pt, \qt\right)
             \frac{z}{z_1} \right] 
  {\cal Q}\left(\frac{x}{z},|\qt+\pt'|^2\right) \Bigg].
\end{split}
\label{eq:final}
\end{equation}
From the above expression we can see that the $\frac{1}{\pt'^2}$ singularity
of the quark term is now regularized by the Regge formfactor
$\Delta_R(z,\qt^2,\mu^2)\equiv\exp\left(-\overline\alpha_s\ln\frac{1}{z}\ln\frac{\qt^2}{\mu^2}\right)$
in a direct analogy with the gluonic term.
We note that the above resummation can be in a straight forward manner
extended~\cite{Kutak:2011fu,Kutak:2012qk,Hentschinski:2016wya} to the situation where the gluon
density is large and therefore subject to a nonlinear evolution equation,
taking into account saturation effects~\cite{Gribov:1984tu}.

\begin{figure}[t]
\centering
\includegraphics[width=0.47\textwidth]{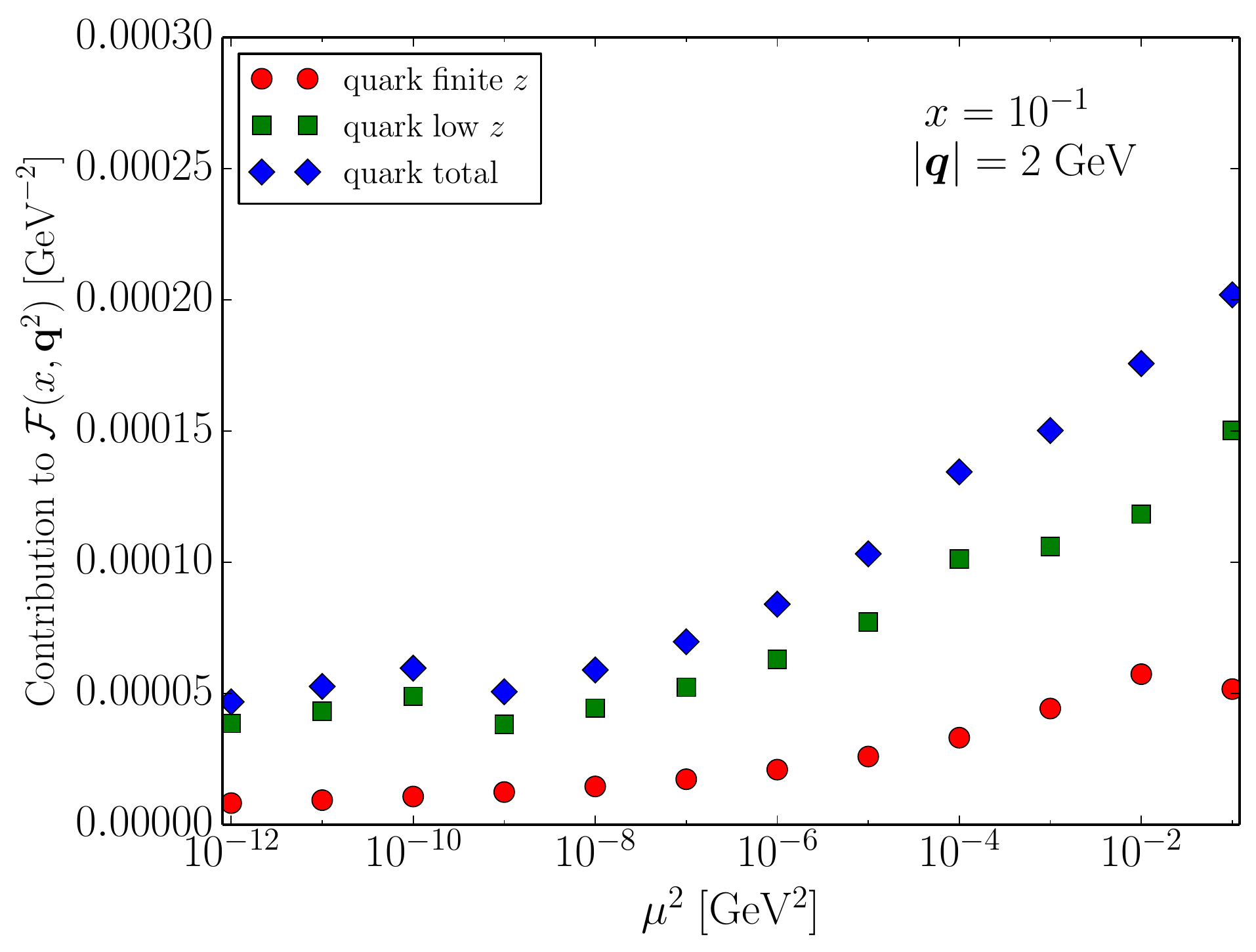}
\hspace{15pt}
\includegraphics[width=0.46\textwidth]{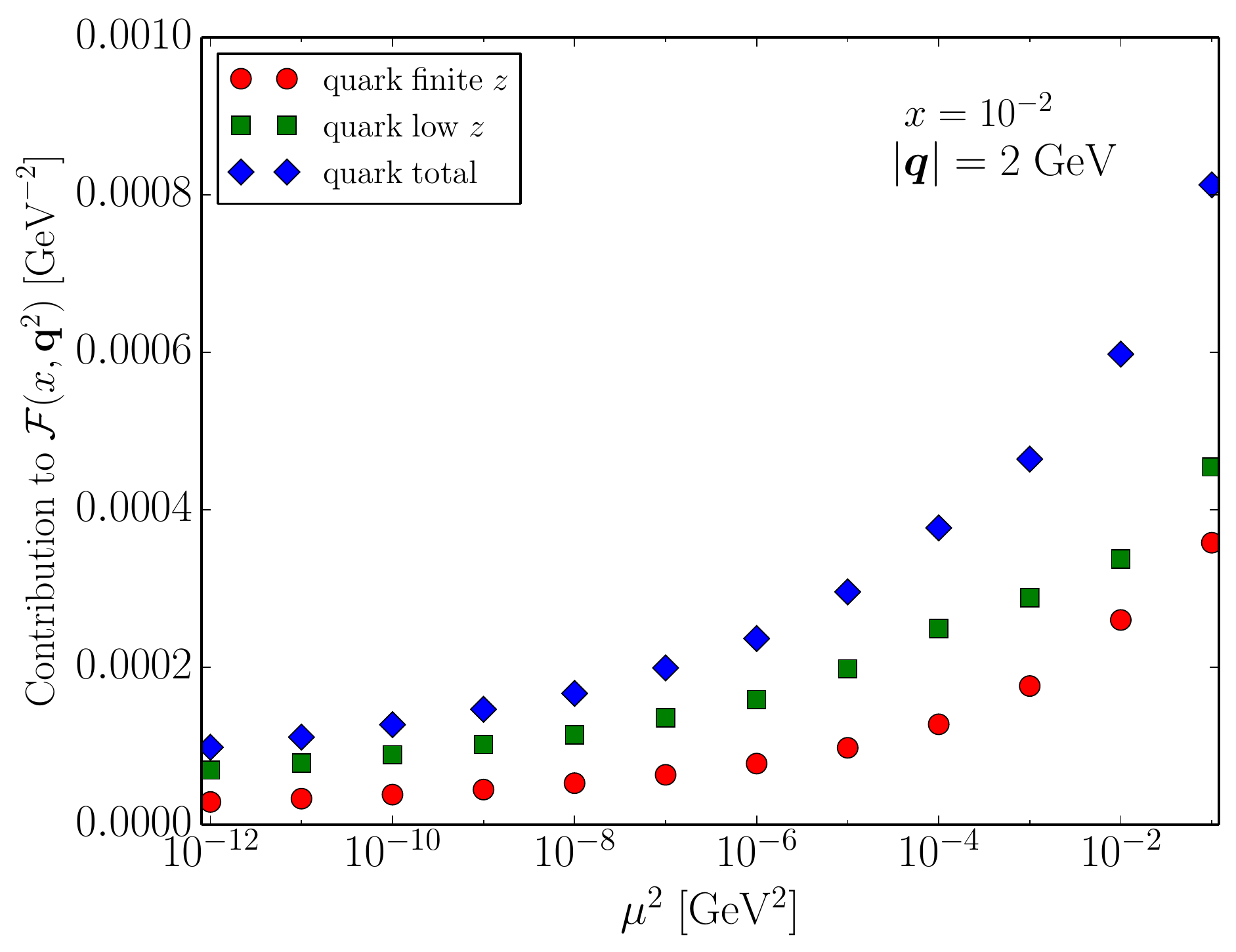}
\caption{The figure visualizes the cutoff dependence of the low $z$ and
finite $z$ quark terms contributing to the gluon density.}
\label{fig:figurequark}
\end{figure}

As a last step we check the stability of the obtained result~\eqref{eq:final}
with respect to the cut-off $\mu$. To this end we perform a convolution of the
low $z$ and fine $z$ parts of the $P_{gq}$ kernel with the quark density~\cite{Kutak:2016mik}.
This leads us to the conclusion that as $\mu^2\to0$ the cutoff dependence gets weaker,
see Fig.~\ref{fig:figurequark}.

%%%%%%%%%%%%%%%%%%%%%%%%
\section{Conclusions}
%%%%%%%%%%%%%%%%%%%%%%%%
We have generalized a framework of Catani and Hautmann and used it to calculate the real emission
$k_\perp$-dependent $P_{qq}$, $P_{gq}$ and $P_{qg}$ splitting functions. These splitting functions
were then used to construct evolution equation for gluons, receiving contribution from quarks.
We have demonstrated that the singularity of the $P_{gq}$ kernel can be regularized by the means
of the same cut-off as in case of BFKL kernel. Furthermore, resumming the combined contribution
of the virtual part and the small $\pt'$ real part of the BFKL kernel to all orders in the strong
coupling, one finds that both the pure gluonic contribution to the evolution equation as well as
the quark induced term are finite if we send this cut-off to zero.
In particular we have demonstrated, via performing one iteration
of the kernels, that the equation has a realistic chance to be stable against
variation of the cutoff parameter, since after one iteration the result
stabilizes. 

To perform a fully consistent study of the complete system of
$k_T$-dependent evolution equations,  we need to calculate the virtual
contributions to the quark-to-quark splitting function in the framework of
$k_T$-factorization. 
Additionally, to demonstrate the completeness of our framework, the $P_{gg}$ splitting
function should be recalculated using the presented approach.

%====================================================================
%====================================================================
\bibliographystyle{utphys_spires}
\bibliography{refs}
% \begin{thebibliography}{99}
%  \bibitem{...} ....
% \end{thebibliography}

%====================================================================
\end{document}